\documentclass[conference,a4paper]{IEEEtran}%asli

\usepackage{amsmath,epsfig,amssymb,verbatim,amsopn,multirow}
\usepackage{balance}
\usepackage[usenames,dvipsnames]{color}
\usepackage[all]{xy}
\usepackage{url}
\usepackage{amsfonts}
\usepackage{amssymb}
\usepackage{epsfig}
\usepackage{slashbox}
\usepackage{bm}
\usepackage{amsmath,amssymb, amstext}
\usepackage{graphicx}
\usepackage{epstopdf}
\usepackage{hyperref}
\usepackage[english]{babel}
\usepackage{graphicx}
\usepackage{caption}
\usepackage{enumerate}
\usepackage{epsfig}
\usepackage{bm}
\usepackage{amsmath,amssymb, amstext}
\usepackage{graphicx}
\usepackage{epstopdf}
\usepackage{amsmath,epsfig,amssymb,verbatim,amsopn,cite,subfigure,multirow}
\usepackage{thmtools}
\usepackage{mathtools}
\usepackage{balance}
\usepackage[usenames,dvipsnames]{color}
\usepackage[all]{xy}
\usepackage{url}
\usepackage{amsfonts}
\usepackage{amssymb}
\usepackage{color}
\usepackage{bm}
\usepackage{epsfig}
\usepackage{amsmath,amssymb, amstext}
\usepackage{graphicx}
\usepackage{epstopdf}
\usepackage{algorithm}
\usepackage{algorithmic}
\usepackage{subfigure}
\usepackage[subfigure]{tocloft}
\usepackage[mathscr]{euscript}
\usepackage{multirow}
\usepackage{cleveref}
\usepackage[font=small,labelfont=bf,labelsep=space]{caption}
  {\proof}{\proofend}

\newcounter{mytempeqcounter}

\newcommand{\superscript}[1]{\ensuremath{^{\textrm{#1}}}}

\definecolor{light-gray}{gray}{0.65}

\begin{document}

\title{Compound Multiple Access Channel with Confidential Messages}

\author{\IEEEauthorblockN{ Hassan Zivari-Fard\superscript{\dag,}\superscript{\dag\dag}, Bahareh Akhbari\superscript{\dag\dag}, Mahmoud Ahmadian-Attari\superscript{\dag\dag}, Mohammad Reza Aref\superscript{\dag}}
\IEEEauthorblockA{
\superscript{\dag}Information Systems and Security Lab (ISSL), Sharif University of Technology, Tehran, Iran\\
Email: {{hassan\_zivari}@ee.kntu.ac.ir, {aref}@sharif.edu}\\
\superscript{\dag\dag}Department of ECE, K. N. Toosi University of Technology, Tehran, Iran\\
Email: {\{akhbari, mahmoud\}@eetd.kntu.ac.ir}}}
\maketitle
%%%%%%%%%%%%%%%%%%%%%%%%%%%%%%%%%%%%%%%%%%%%%%%%%%%% ABSTRACT %%%%%%%%%%%%%%%%%%%%%%%%%%%%%%%%%%%%%%%
\footnote{This work was partially supported by Iran Telecom Research Center under contract no. 17175/500.}
\begin{abstract}
In this paper, we study the problem of secret communication over a Compound Multiple Access Channel (MAC). In this channel, we assume that one of the transmitted messages is confidential that is only decoded by its corresponding receiver and kept secret from the other receiver. For this proposed setting (compound MAC with confidential messages), we derive general inner and outer bounds on the secrecy capacity region. Also, as examples, we investigate 'Less noisy' and 'Gaussian' versions of this channel, and extend the results of the discrete memoryless version to these cases. Moreover, providing numerical examples for the Gaussian case, we illustrate the comparison between achievable rate regions of compound MAC and compound MAC with confidential messages.
\end{abstract}
\IEEEpeerreviewmaketitle
%\vspace{-4mm}
%%%%%%%%%%%%%%%%%%%%%%%%%%%%%%%%%%%%%%%%%%%%%%%%%%% INTRODUCTION %%%%%%%%%%%%%%%%%%%%%%%%%%%%%%%%%%%%
\section{INTRODUCTION}
\fontsize{10}{11}
\selectfont
The wire-tap channel was first introduced by Wyner in 1975 \cite{Wyner}. His model consisted of a transmitter, a receiver and an eavesdropper. In the Wyner model, the eavesdropper channel was a degraded version of the legitimate receiver channel. Csisz\'{a}r and K\"{o}rner extended the wire-tap channel to a more generalized model called the broadcast channel with confidential message \cite{CsiszarKorner}. More recently, sending a confidential message over multiple-user channels have been studied under various different models \cite{cooperation, LiangPoor, EkremUlukus, Yassaee, ISIT2012, Maric}. We also refer the reader to \cite{ITSliang} for a recent survey of the research progress in this area. The discrete memoryless compound Multiple Access Channel (MAC), Gaussian compound MAC with a common message and conferencing decoders, and also the compound MAC when both encoders and decoders cooperate via conferencing links were considered in \cite{simeone}.

In \cite{cooperation} the authors studied the effect of users' cooperation in multiple access channel when transmitting a confidential message. There, active cooperation between two trusted users is attained through a generalized feedback channel. In \cite{LiangPoor} a discrete memoryless multiple access channel with confidential messages was studied where each user uses the output of generalized feedback to eavesdrop the other user's private message. Ekrem and Ulukus derived n-letter inner and outer bounds for the multiple access wire-tap channel with no common message \cite{EkremUlukus}. In \cite{ISIT2012}, the authors studied this model assuming that there exists a common message and that the eavesdropper is unable to decode it. They also derived a rate region under the strong secrecy criterion.

\begin{figure}%*}[ht]
  \centering
  \includegraphics[width=8.87cm]{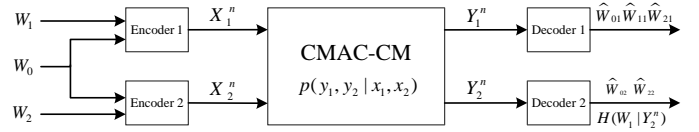}% asli 12.50cm 8.50
  \caption{\small{Compound Multiple Access Channel with Confidential Messages (CMAC-CM)}}
  \label{fig}
\end{figure}

In this paper, we consider Compound Multiple Access Channel with Confidential Messages (CMAC-CM). Actually, in wireless networks, there may be a scenario in which some of the users have confidential information that wish to be kept secret from illegal users. In fact, in terms of information the users can be divided into legitimate and illegal users. Legitimate users are allowed to decode \emph{all} the transmitted information (including common and private messages of all the transmitters), while illegal users are allowed to decode only the messages of their intended transmitters. Motivated by this scenario, we consider CMAC-CM as a building block of this setting. In this model, while each of the transmitters sends its own private message, both of them have a common message. One of the transmitters' private message ($W_1$) is confidential and only decoded by the first receiver and kept secret from the second receiver. The common message $W_0$ and private message $W_2$ are decoded by both receivers (see Fig. \ref{fig}). For this model we derive single letter inner and outer bounds on the secrecy capacity region. We also consider two examples for this channel: Less noisy and Gaussian CMAC-CM.

This paper is organized as follows. In Section II, the system model is described. In Section III, an outer bound on the secrecy capacity region of CMAC-CM and also an achievable secrecy rate region for CMAC-CM are derived. Two examples are given in Section IV. The paper is concluded in Section V.
%*}
\setlength{\textfloatsep}{10pt plus 1.0pt minus 2.0pt}

\section{SYSTEM MODEL}
Consider a discrete memoryless CMAC-CM with four-terminals as shown in Fig. \ref{fig}. The finite sets $\mathcal{X}_1$,$\mathcal{X}_2$,$\mathcal Y_{1}$,$\mathcal Y_{2}$ and the transition probability distribution
$p(y_{1},y_{2}|{x_1},{x_2})$ are the constitutive components of this channel.
Here, $X_{1}$ and $X_{2}$ are the channel inputs from the transmitters. Also $Y_{1}$ and $Y_{2}$ are the channel outputs at the receiver~1 and receiver~2, respectively.
Throughout this paper, the random variables are denoted by capital letters e.g., X, Y, and their realizations by lower case letters e.g. x, y.
The set of $\varepsilon -$strongly jointly typical sequences of length $n$, on joint distribution $p(x,y)$ is denoted by
$A_\varepsilon ^n({P_{X,Y}})$.  We use $X_i^n$, to indicate vector $(X_{i,1},X_{i,2},\ldots,X_{i,n})$, and $X_{i,j}^k$ to indicate vector $(X_{i,j},X_{i,j+1},\ldots,X_{i,k})$.
Before discussing the achievability rate, we first define a code for the channel as follows.

\textbf{Definition 1:} A  $({M_0},{M_1},{M_2},n,P_e^n)$ code for the CMAC-CM (Fig. \ref{fig}) consists of the following:
i) Two message sets $({W_0},{W_1})$ and $({W_0},{W_2})$ that are uniformly distributed over $[1:{M_0}] \times [1:{M_1}]$ and $[1:{M_0}] \times [1:{M_2}]$, respectively, where messages $W_u \in {\mathcal{W}}_u = \{ 1,2,...,{M_u}\}$ and $u=0,1,2$. Note that ${W_0}$, ${W_1}$ and ${W_2}$ are independent.
ii) A stochastic encoder $f$ for transmitter 1 is specified by the matrix of conditional probability $f(X_1^n|{w_0},{w_1})$, where $X_1^n \in \mathcal{X}_1^n$, $w_0\in\mathcal{W}_0$, $w_1\in\mathcal{W}_1$ are channel input, common and private message sets respectively, and $\sum\nolimits_{X_1^n} {f(X_1^n|{w_0},{w_1})}  = 1$. Note that $f(X_1^n|{w_0},{w_1})$ is the probability of encoding message pair $({w_0},{w_1})$ to the channel input $X_1^n$.
iii) A deterministic encoder $g$ for transmitter 2 which is the mapping $ g:\mathcal{W}_0 \times \mathcal{W}_2 \to \mathcal{X}_2^n$ for generating codewords $X_2^n = g({w_0},{w_2})$.
%\end{enumerate}%ta inja item 2
iv) A decoding function $\phi :\mathcal Y_{1}^n \to \mathcal{W}_0 \times \mathcal{W}_1 \times \mathcal{W}_2$ at the receiver 1 that assigns $({\widehat W_{01}},\widehat W_{11},\widehat W_{21}) \in [1:{M_0}] \times [1:{M_1}] \times [1:{M_2}]$ to received sequence $y_{1}^n$. v) A decoding function $\rho :{\mathcal Y_2^n} \to \mathcal{W}_0 \times \mathcal{W}_2$, at the receiver 2 that assigns $(\widehat W_{02},\widehat W_{22})\in [1:M_0]\times [1:{M_2}]$ to received sequence $y_{2}^n$.
The probability of error is defined as,
\begin{align}
\label{error}
P_e^n =\mbox{Pr}(&\widehat W_{0j} \ne {W_0} \,\,\,\, \mbox{for} \,\,\,\, j=1,2 \,\,\,\, \mbox{or} \nonumber\\
&\widehat W_{11} \ne {W_1} \,\,\,\, \mbox{or} \,\,\,\, \widehat W_{2j} \ne {W_2} \,\,\,\, \mbox{for} \,\,\,\, j=1,2).
\end{align}

The ignorance level of receiver 2 with respect to the confidential message is measured by the normalized equivocation $\frac{1}{n}H({W_1}|{Y_{2}^n})$.

\textbf{Definition 2:} A rate tuple $({R_0},{R_1},{R_2})$ is said to be achievable for CMAC-CM, if for any $\delta > 0$ there exists a $({M_0},{M_1},{M_2},n,P_e^n)$ code as
\begin{align}
P_e^n &< \varepsilon \\
{M_0} &\ge {2^{n{R_0}}},{M_1} \ge {2^{n{R_1}}},{M_2} \ge {2^{n{R_2}}}\\
\label{Tarif Secrecy}
{R_1} &- \frac{1}{n}H({W_1}|{Y_{2}^n}) \le \delta
\end{align}

\section{MAIN RESULTS}
\subsection{Outer Bound}
\textbf{Theorem 1:} (Outer bound) The secrecy capacity region for the CMAC-CM is included in the set of rates satisfying
\begin{align}
    \label{R_0}
    {R_0} &\le \min \{I(U;Y_{1}),I(U;Y_{2})\}  \\
    \label{R_1}
    {R_1} &\le I(V_{1};Y_{1}|U,V_{2}) - I(V_{1};Y_{2}|U,V_{2})  \\
    \label{R_2}
    {R_2} &\le \min \{I(V_{2};Y_{1}),I(V_{2};Y_{2})\} \\
    \label{{R_1} + {R_2}}
    {R_1+R_2} &\le I({V_{1}},V_{2};Y_{1}) - I(V_{1};Y_{2}|U,V_{2})
\end{align}
for some joint distribution
\begin{equation}
p(u)p(v_{1},v_{2}|u)p(x_{1}|v_{1})p(x_{2}|v_{2})p(y_{1},y_{2}|x_{1},x_{2}).\label{tozi}
\end{equation}

\textbf{Remark 1:} If we set $W_{2}=\emptyset$ and thus $R_2=0$ in Theorem~1, the region reduces to the region of the broadcast channel with confidential messages discussed in \cite{CsiszarKorner} by Csisz\'{a}r and K\"{o}rner.

\textbf{Proof (Theorem 1)}:
We next show that any achievable rate tuples satisfies (\ref{R_0})-(\ref{{R_1} + {R_2}})
for some distribution factorized as (\ref{tozi}).
Consider a code $(M_0,M_1,M_2,n,P_e^n)$ for the CMAC-CM. Applying Fano's inequality \cite{ElgamalKim} results in
\begin{align}
   &H(W_0,W_1,W_2|Y_{1}^n)\leq n{\varepsilon_1}\\
   &H(W_0,W_2|Y_{2}^n)\leq n{\varepsilon_2}
\end{align}

We first derive the bound on $R_1$. Note that the perfect secrecy (\ref{Tarif Secrecy}) implies that
\begin{equation}\label{security}
  n{R_1} - n\delta  \le H({W_1}|{Y_{2}^n}).
\end{equation}

Hence, we derive the bound on $H({W_1}|{Y_{2}^n})$ as following:
\begin{align}
H({W_1}|{Y_{2}^n}) &= H({W_1}|{Y_{2}^n},{W_0},{W_2})+ I({W_1};{W_0},{W_2}|{Y_{2}^n})\nonumber\\
&= H({W_1}|{Y_{2}^n},{W_0},{W_2})+H({W_0},{W_2}|{Y_{2}^n}) \nonumber\\
&\,\,\,\,\,\,- H({W_0},{W_2}|{Y_{2}^n},{W_1})\nonumber\\
&\le H({W_1}|{Y_{2}^n},{W_0},{W_2})+n{\varepsilon_2}\nonumber\\
&\le H({W_1}|{Y_{2}^n},{W_0},{W_2}) - H({W_1}|{Y_{1}^n},{W_0},{W_2}) \nonumber\\
\label{ghabli}
&\,\,\,\,\,\,+ n{\varepsilon_1}+n{\varepsilon_2}
\end{align}where the first and the second inequality are due to Fano's inequalities. Now, based on (\ref{ghabli}) we have
{\small
\begin{align}
H({W_1}|{Y_{2}^n}) &\le I({W_1};{Y_{1}^n}|{W_0},{W_2}) \nonumber\\
&\,\,\,\,\,\,- I({W_1};{Y_{2}^n}|{W_0},{W_2}) + n\varepsilon \nonumber\\
&= \sum\limits_{i = 1}^n {[I({W_1};{Y_{1,i}}|{Y_{1}^{i - 1}},{W_0},{W_2})}\nonumber\\
&\,\,\,\,\,\,- I({W_1};{Y_{2,i}}|Y_{2,i + 1}^n,{W_0},{W_2})]  + n\varepsilon \nonumber\\
&= \sum\limits_{i = 1}^n {[I({W_1},Y_{2,i + 1}^n;{Y_{1,i}}|{Y_{1}^{i - 1}},{W_0},{W_2})}\nonumber\\
&\,\,\,\,\,\,- I(Y_{2,i + 1}^n;{Y_{1,i}}|{Y_{1}^{i - 1}},{W_0},{W_1},{W_2}) \nonumber\\
&\,\,\,\,\,\,-I({W_1},{Y_{1}^{i - 1}};{Y_{2,i}}|Y_{2,i + 1}^n,{W_0},{W_2})\nonumber\\
&\,\,\,\,\,\,+ I({Y_{1}^{i - 1}};{Y_{2,i}}|Y_{2,i + 1}^n,{W_0},{W_1},{W_2})] + n\varepsilon \nonumber\\
&= \sum\limits_{i = 1}^n {[I(Y_{2,i + 1}^n;{Y_{1,i}}|{Y_{1}^{i - 1}},{W_0},{W_2})}\nonumber\\
&\,\,\,\,\,\,+ I({W_1};{Y_{1,i}}|Y_{2,i + 1}^n,{Y_{1}^{i - 1}},{W_0},{W_2}) \nonumber\\
&\,\,\,\,\,\,-I(Y_{2,i + 1}^n;{Y_{1,i}}|{Y_{1}^{i - 1}},{W_0},{W_1},{W_2})\nonumber\\
&\,\,\,\,\,\,-I({Y_{1}^{i - 1}};{Y_{2,i}}|Y_{2,i + 1}^n,{W_0},{W_2})\nonumber\\
&\,\,\,\,\,\,-I({W_1};{Y_{2,i}}|Y_{2,i + 1}^n,{Y_{1}^{i - 1}},{W_0},{W_2})\nonumber\\
&\,\,\,\,\,\,+I({Y_{1}^{i - 1}};{Y_{2,i}}|Y_{2,i + 1}^n,{W_0},{W_1},{W_2})] + n\varepsilon \nonumber\\
&= \sum\limits_{i = 1}^n {[I({W_1};{Y_{1,i}}|Y_{2,i + 1}^n,{Y_{1}^{i - 1}},{W_0},{W_2})}\nonumber\\
&\,\,\,\,\,\,- I({W_1};{Y_{2,i}}|Y_{2,i + 1}^n,{Y_{1}^{i - 1}},{W_0},{W_2})]  + n\varepsilon\nonumber
\end{align}}where $\varepsilon  = {\varepsilon _1} + {\varepsilon _2}$. The last equality is due to \cite[Lemma 7]{CsiszarKorner} where:
{\small
\begin{align}
\label{tasavie ciszar 1}
\sum\limits_{i = 1}^n {I(Y_{2,i + 1}^n;{Y_{1,i}}|{Y_{1}^{i - 1}},{W_0},{W_2})}=\nonumber\\
\sum\limits_{i = 1}^n {I({Y_{1}^{i - 1}};{Y_{2,i}}|Y_{2,i + 1}^n,{W_0},{W_2})}
\end{align}} and
{\small
\begin{align}
\label{tasavie ciszar 2}
\sum\limits_{i = 1}^n {I(Y_{2,i + 1}^n;{Y_{1,i}}|{Y_{1}^{i - 1}},{W_0},{W_1},{W_2})}=\nonumber\\
\sum\limits_{i = 1}^n {I({Y_{1}^{i - 1}};{Y_{2,i}}|Y_{2,i + 1}^n,{W_0},{W_1},{W_2})}.
\end{align}}

So, we have
{\small
\begin{align}
H(&{W_1}|{Y_{2}^n}) \le \sum\limits_{i = 1}^n {[I({W_1};{Y_{1,i}}|U_{i},V_{2,i})}\nonumber \\
&\,\,\,\,\,\,\,\,\,\,\,\,\,\,\,\,\,\,\,\,\,\,\,\,\,\,\,\,\,\,\,\,\,\,\,\,\,\,\,\,\,
- I({W_1};{Y_{2,i}}|U_{i},V_{2,i})]+n\varepsilon \nonumber \\
&= \sum\limits_{i = 1}^n {[I(V_{1,i};{Y_{1,i}}|U_{i},V_{2,i})}- I(V_{1,i};{Y_{2,i}}|U_{i},V_{2,i})]+ n\varepsilon
\end{align}}where the equalities resulting from the following definitions of the random variables
{\small
\begin{align}
\label{U_i}
&U_{i} = Y_{2,i + 1}^n,{Y_{1}^{i - 1}},{W_0}\\
\label{V_{1,i}}
&V_{1,i} = ({U_i},{W_1})\\
\label{V_{2,i}}
&V_{2,i} = ({U_i},{W_2}).
\end{align}}

Now, we have
{\small
\begin{align}
H({W_1}|{Y_{2}^n}) &\le n\sum\limits_{i = 1}^n {\frac{1}{n}[I(V_{1,Q};{Y_{1,Q}}|{U_Q},V_{2,Q},Q = i)}\nonumber\\
&\,\,\,\,\,\,- I(V_{1,Q};{Y_{2,Q}}|{U_Q},V_{2,Q},Q = i)] + n\varepsilon \nonumber\\
&= n\sum\limits_{i = 1}^n {p(Q = i)[I(V_{1,Q};{Y_{1,Q}}|{U_Q},V_{2,Q},Q = i)}\nonumber\\
&\,\,\,\,\,\,- I(V_{1,Q};{Y_{2,Q}}|{U_Q},V_{2,Q},Q = i)]  + n\varepsilon \nonumber\\
&= n[I(V_{1,Q};{Y_{1,Q}}|{U_Q},V_{2,Q},Q) \nonumber\\
&\,\,\,\,\,\,- I(V_{1,Q};{Y_{2,Q}}|{U_Q},V_{2,Q},Q)] + n\varepsilon \nonumber\\
\label{Nerkh R_1}
&= n[I(V_{1};Y_{1}|U,V_{2}) - I(V_{1};Y_{2}|U,V_{2})] + n\varepsilon
\end{align}}where $V_{1,Q}=V_1$, $V_{2,Q}=V_2$, $Y_{1,Q}=Y_{1}$, $Y_{2,Q}=Y_{2}$, $(U_{Q},Q)=U$ and $Q$ has a uniform distribution over $\{1,2,...,n\}$ outcomes. Now, we derive the bound on ${R_2}$ as following:
{\small
\begin{align}
n{R_2} &= H({W_2}) = H({W_2}|{W_0}) \nonumber \\
&= I({W_2};{Y_{1}^n}|{W_0}) + H({W_2}|{Y_{1}^n},{W_0})\nonumber \\
\label{20000}
&\le I({W_2};{Y_{1}^n}|{W_0}) + n{\varepsilon_1}
\end{align}}where the second equality results from independence of ${W_0},{W_1},{W_2}$, and the inequality is due to the Fano's inequality. Now, based on (\ref{20000})
{\small
\begin{align}
n{R_2} &\le \sum\limits_{i = 1}^n {I(W_2;{Y_{1,i}}|{Y_{1}^{i - 1}},{W_0})}  + n{\varepsilon_1} \nonumber \\
&\le \sum\limits_{i = 1}^n {I({W_2},Y_{2,i + 1}^n;{Y_{1,i}}|{Y_{1}^{i - 1}},{W_0})}  + n{\varepsilon_1} \nonumber \\
&\le \sum\limits_{i = 1}^n {I({W_2},{W_0},Y_{2,i + 1}^n,Y_{1}^{i - 1};{Y_{1,i}})} + n{\varepsilon_1} \nonumber \\
&= \sum\limits_{i = 1}^n {I(V_{2,i},{U_i};{Y_{1,i}})} + n{\varepsilon_1}
\end{align}}where the last equality follows from (\ref{U_i}) and (\ref{V_{2,i}}). Now, we have
{\small
\begin{align}
{R_2} &\le \sum\limits_{i = 1}^n {\frac{1}{n}[I(V_{2,i},{U_i};{Y_{1,i}})]}  + n{\varepsilon_1} \nonumber \\
&= \sum\limits_{i = 1}^n {\frac{1}{n}[I(V_{2,Q},{U_Q};{Y_{1,Q}}|Q=i)]}  + n{\varepsilon_1} \nonumber \\
&= \sum\limits_{i = 1}^n {p(Q = i)[I(V_{2,Q},U_{Q};{Y_{1,Q}}|Q=i)]}  + n{\varepsilon_1} \nonumber \\
&= I(V_{2,Q},U_{Q};Y_{1,Q}|Q) + n{\varepsilon_1} \nonumber \\
&= I(V_{2,Q},U_{Q},Q;Y_{1,Q}) + n{\varepsilon_1} \nonumber \\
&= I(V_{2,Q};{Y_{1,Q}}) + I(U_{Q};{Y_{1,Q}}|V_{2,Q})\nonumber\\
&\,\,\,\,\,\,+ I(Q;{Y_{1,Q}}|V_{2,Q},U_{Q}) + n{\varepsilon_1} \nonumber \\
&= I(V_{2,Q};{Y_{1,Q}})+ n{\varepsilon_1} \nonumber \\
\label{Nerkh R_2Y}
&= I(V_{2};Y_{1}) + n{\varepsilon_1}
\end{align}}where fourth equality result from independence of $Q$ and $Y_Q$. Sixth equality is due to $I(U_{Q};{Y_{1,Q}}|V_{2,Q})=0$ (see (\ref{U_i}) and (\ref{V_{2,i}})), and $I(Q;{Y_{1,Q}}|V_{2,Q},U_{Q})=0$ since $Q$ is independent of $Y_{1,Q}$. The last equality results by setting $V_{2,Q}=V_{2}$, $Y_{1,Q}=Y_{1}$.

On the other hand, we have:
{\small
\begin{align}
n{R_2} &= H({W_2}) = H({W_2}|{W_0}) \nonumber \\
&= I({W_2};{Y_{2}^n}|{W_0}) + H({W_2}|{Y_{2}^n},{W_0})\nonumber \\
&\le I({W_2};{Y_{2}^n}|{W_0}) + n{\varepsilon_2}
\end{align}}where the inequality is due to the Fano's inequality. Next, we have
{\small
\begin{align}
n{R_2} &\le \sum\limits_{i = 1}^n {I(W_2;{Y_{2,i}}|{Y_{2,i + 1}^n},{W_0})}  + n{\varepsilon_2} \nonumber \\
&\le \sum\limits_{i = 1}^n {I({W_2},{Y_{1}^{i - 1}};{Y_{2,i}}|{Y_{2,i + 1}^n},{W_0})}  + n{\varepsilon_2} \nonumber \\
&\le \sum\limits_{i = 1}^n {I({W_2},Y_{2,i + 1}^n,Y_{1}^{i - 1},{W_0};{Y_{2,i}})}  + n{\varepsilon_2} \nonumber \\
&= \sum\limits_{i = 1}^n I(U_{i},V_{2,i};{Y_{2,i}}) + n{\varepsilon_2}
\end{align}}where the last equality follows from (\ref{U_i}) and (\ref{V_{2,i}}). Now, by applying the same time-sharing strategy as (\ref{Nerkh R_2Y}) we have
{\small
\begin{equation}\label{Nerkh R_2Z}
{R_2} \leq I(V_{2};Y_{2}).
\end{equation}}

Now, we derive the bound on $n({R_1 + R_2})$ as following:
{\small
\begin{align}
n({R_1} + {R_2}) &= H({W_1},{W_2}) = H({W_1},{W_2}|{W_0})\nonumber \\
&= I({W_1},{W_2};{Y_{1}^n}|{W_0}) + H({W_1},{W_2}|{Y_{1}^n},{W_0})\nonumber \\
&\le I({W_1},{W_2};{Y_{1}^n}|{W_0}) + n{\varepsilon_1}\nonumber
\end{align}}where the inequality is due to the Fano's inequality. Now, we have
{\small
\begin{align}
n({R_1} + {R_2}) &\leq I({W_1},{W_2};{Y_{1}^n}|{W_0})\nonumber \\
&\,\,\,\,\,\,- (H({W_1}) - H({W_1}|{Y_{2}^n}) - n\delta ) + n{\varepsilon_1}\nonumber \\
&= I({W_1},{W_2};{Y_{1}^n}|{W_0}) - H({W_1}|{W_0},{W_2}) \nonumber \\
&\,\,\,\,\,\,+ H({W_1}|{Y_{2}^n}) + n({\varepsilon_1}+\delta)\nonumber\\
&= I({W_1},{W_2};{Y_{1}^n}|{W_0}) - H({W_1}|{W_0},{W_2})\nonumber \\
&\,\,\,\,\,\,+ H({W_1}|{Y_{2}^n},{W_0},{W_2}) \nonumber \\
&\,\,\,\,\,\,+ I({W_1};{W_0},{W_2}|{Y_{2}^n}) + n({\varepsilon_1}+\delta) \nonumber \\
&\leq I({W_1},{W_2};{Y_{1}^n}|{W_0}) \nonumber \\
&\,\,\,\,\,\,- I({W_1};{Y_{2}^n}|{W_0},{W_2}) + n{\varepsilon_2}+n({\varepsilon_1}+\delta) \nonumber
\end{align}}where the first inequality is a consequence of (\ref{Tarif Secrecy}) and the last inequality is due to the Fano's inequality. We also define $\varepsilon_3  = {\varepsilon _1} + {\varepsilon_2} + \delta$. So, we have
{\small
\begin{align}
n({R_1} + {R_2}) &\leq \sum\limits_{i = 1}^n {[I({W_1},{W_2};{Y_{1,i}}|{Y_{1}^{i - 1}},{W_0})} \nonumber \\
&\,\,\,\,\,\,- I({W_1};{Y_{2,i}}|Y_{2,i + 1}^n,{W_0},{W_2})]  + n{\varepsilon_3} \nonumber\\
&= \sum\limits_{i = 1}^n {[I({W_1},{W_2},Y_{2,i + 1}^n;{Y_{1,i}}|{Y_{1}^{i - 1}},{W_0})}\nonumber \\
&\,\,\,\,\,\,-  I(Y_{2,i + 1}^n;{Y_{1,i}}|{Y_{1}^{i - 1}},{W_0},{W_1},{W_2})\nonumber\\
&\,\,\,\,\,\,- I({W_1},{Y_{1}^{i - 1}};{Y_{2,i}}|Y_{2,i + 1}^n,{W_0},{W_2}) \nonumber \\
&\,\,\,\,\,\,+ I({Y_{1}^{i - 1}};{Y_{2,i}}|Y_{2,i + 1}^n,{W_0},{W_1},{W_2})] + n{\varepsilon_3} \nonumber\\
&= \sum\limits_{i = 1}^n {[I({W_1},{W_2},Y_{2,i + 1}^n;{Y_{1,i}}|{Y_{1}^{i - 1}},{W_0})} \nonumber \\
&\,\,\,\,\,\,- I({W_1},{Y_{1}^{i - 1}};{Y_{2,i}}|Y_{2,i + 1}^n,{W_0},{W_2})] + n{\varepsilon_3}\nonumber
\end{align}}where the last equality is a consequence of (\ref{tasavie ciszar 2}). Hence, we have
{\small
\begin{align}
n({R_1} + {R_2}) &\leq \sum\limits_{i = 1}^n {[I({W_1},{W_2},Y_{2,i + 1}^n;{Y_{1,i}}|{Y_{1}^{i - 1}},{W_0})}\nonumber\\
&\,\,\,\,\,\,- I({W_1};{Y_{2,i}}|Y_{2,i + 1}^n,{Y_{1}^{i - 1}},{W_0},{W_2})] + n{\varepsilon_3} \nonumber\\
&\le \sum\limits_{i = 1}^n {[I({W_0},{W_1},{W_2},{Y_{1}^{i - 1}},Y_{2,i + 1}^n;{Y_{1,i}})} \nonumber\\
&\,\,\,\,\,\,- I({W_1};{Y_{2,i}}|Y_{2,i + 1}^n,{Y_{1}^{i - 1}},{W_0},{W_2})] + n{\varepsilon_3}.\nonumber%\\
\end{align}}

Now, we have
{\small
\begin{align}
n({R_1} + {R_2}) &\leq \sum\limits_{i = 1}^n {[I({U_i},V_{1,i},V_{2,i};{Y_{1,i}})} \nonumber\\
&\,\,\,\,\,\,- I(V_{1,i};{Y_{2,i}}|{U_i},V_{2,i})]  + n{\varepsilon_3} \nonumber
\end{align}}where the inequality follows from (\ref{U_i})-(\ref{V_{2,i}}). Now, by applying the same time-sharing strategy as before, we have:
{\small
\begin{equation}\label{Nerkh R_1 + R_1}
{R_1} + {R_2} \leq I({V_{1}},V_{2};Y_{1}) - I(V_{1};Y_{2}|U,V_{2}).
\end{equation}}

Now, we derive the bound on ${R_0}$ as following:
{\small
\begin{align}
n{R_0} &= H({W_0})= I(W_{0};{Y_{1}^n}) + H(W_{0}|{Y_{1}^n})\nonumber \\
&\le I({W_0};{Y_{1}^n}) + n{\varepsilon_1} \nonumber \\
&= \sum\limits_{i = 1}^n {I({W_0};{Y_{1,i}}|Y_{1}^{i - 1})}  + n{\varepsilon_1} \nonumber \\
&= \sum\limits_{i = 1}^n {[I({W_0},Y_{1}^{i - 1};{Y_{1,i}}) - I(Y_{1}^{i - 1};{Y_{1,i}})]}  + n{\varepsilon_1}. \nonumber
\end{align}}

So, we have
{\small
\begin{align}
n{R_0}&\le \sum\limits_{i = 1}^n {I({W_0},{Y_{1}^{i - 1}};{Y_{1,i}})}  + n{\varepsilon_1} \nonumber \\
&\le \sum\limits_{i = 1}^n {[I({W_0},{Y_{1}^{i - 1}},Y_{2,i + 1}^n;{Y_{1,i}})}\nonumber \\
&\,\,\,\,\,\,- I(Y_{2,i + 1}^n;{Y_{1,i}}|{W_0},{Y_{1}^{i - 1}})] + n{\varepsilon_1} \nonumber \\
&\le \sum\limits_{i = 1}^n {I({W_0},{Y_{1}^{i - 1}},Y_{2,i + 1}^n;{Y_{1,i}})}  + n{\varepsilon_1} \nonumber \\
&= \sum\limits_{i = 1}^n {I({U_i};{Y_{1,i}})}  + n{\varepsilon_1}. \nonumber
\end{align}}

Now, by applying the same time-sharing strategy as before, we have
{\small
\begin{equation}\label{111}
{R_0} \leq {I(U;Y_{1})} + {\varepsilon_1}.
\end{equation}}

Similarly
{\small
\begin{equation}\label{222}
{R_0} \leq {I(U;Y_{2})} + {\varepsilon_2}.
\end{equation}}

Therefore
{\small
\begin{equation}\label{333}
{R_0} \leq \min \{I({U};Y_{1}),I({U};Y_{2})\}.
\end{equation}}

Considering (\ref{security}), (\ref{Nerkh R_1}),(\ref{Nerkh R_2Y}),(\ref{Nerkh R_2Z}),(\ref{Nerkh R_1 + R_1}) and (\ref{333}), the region in (\ref{R_0})-(\ref{{R_1} + {R_2}}) is obtained. This completes the proof.\IEEEQED
\subsection{Achievability}
\textbf{Theorem 2:} An inner bound on the secrecy capacity region is given by:
{\small
\begin{equation}
\bigcup \left\{ \begin{array}{l}
R_{0} \ge 0\,\,,{R_1} \ge 0\,\,,{R_2} \ge 0\\
R_{1} \le I({V_1};Y_{1}|{X_2},U)-I({V_1};Y_{2}|{X_2},U)\\
R_{2} \le \min \{I({X_2};Y_{1}|{V_1},U),I({X_2};Y_{2}|U)\}\\
R_{0} + R_{2} \le I(U,{X_2};Y_{2})\\
R_{1} + R_{2} \le I(V_1,{X_2};Y_{1}|U) - I({V_1};Y_{2}|{X_2},U)\\
R_{0} + R_{1} + R_{2} \le I(V_1,{X_2};Y_{1}) - I({V_1};Y_{2}|{X_2},U)
\end{array} \right.
\end{equation}}where the union is taken over all probability distributions of the form
$p(u,v_1,{x_1},{x_2},y_{1},y_{2}) = p(u)p(v_1|u)p({x_1}|v_1)p({x_2}|u)p(y_{1},y_{2}|{x_1},{x_2})$.

\textbf{Remark 2:} If we convert our model to a multiple access channel with correlated sources by setting $Y_{2}=\emptyset$ and $V_1=X_1$ in Theorem 2, the region reduces to the region of the multiple access channel with correlated sources discussed in \cite{SlepianWolf} by Slepian and Wolf.

\textbf{Remark 3:} If we set $X_{2}=\emptyset$ in Theorem 2, the region includes the region of the broadcast channel with confidential messages discussed in \cite{CsiszarKorner} by Csisz\'{a}r and K\"{o}rner.

\textbf{Proof (Theorem 2):}
Fix $p(u),p(v_1|u),p({x_1}|v_1)$ and $p({x_2}|u)$.
\subsubsection{Codebook generation}
\begin{enumerate}[i)]
    \item  Generate ${2^{n{R_0}}}$ codewords ${u^n}$, each is uniformly drawn from the set $A_{\varepsilon} ^n({P_U})$ indexing by ${u^n}({w_0}),\,\,\,{w_0} \in \{ 1,...,{2^{n{R_0}}}\} $.
    \item  For each codeword ${u^n}({w_0})$, generate ${2^{n\tilde R}}$ codewords $v_{1}^n$ each is uniformly drawn from the set $A_{\varepsilon} ^{(n)}({P_{{V_1}|U}})$, where $\tilde R = {R_1} + I(V_1;Y_{2}|{X_2},U)-\varepsilon$. Then, randomly bin the ${2^{n\tilde R}}$ codewords into ${2^{n{R_1}}}$ bins and label them as $v_{1}^n({w_0},{w_1},l)$. Here, ${w_1}$ is the bin number and ${l} \in \mathcal{L} = \{ 1,...,{2^{n(I({V_1};Y_{2}|{X_2},U)-\varepsilon)}}\} $ is the index of codewords in the bin number ${w_1}$.
    \item  For each codeword ${u^n}({w_0})$, generate ${2^{n{R_2}}}$ codewords $x_2^n({w_0},{w_2})$ each is uniformly drawn from the set $A_{\varepsilon}^{(n)}({P_{{X_2}|U}})$, and label them as $x_2^n({w_0},{w_2})$, ${w_2} \in \{ 1,...,{2^{n{R_2}}}\} $.
\end{enumerate}
\subsubsection{Encoding}
To send the message pair $(w_{0},w_{1})$, the encoder $f$ first randomly chooses index $l$ corresponding to $(w_{0},w_{1})$ and then, generates a codeword $X_1^n$ at random according to $\prod\nolimits_{i = 1}^n {p({x_{1,i}}|{v_{1,i}})}$. Transmitter 2 uses the deterministic encoder for sending $({w_0},{w_2})$ and sends codeword $x_2^n({w_0},{w_2})$.
\subsubsection{Decoding and Probability of error}
\begin{itemize}
  \item Receiver 1 declares that the indices of $(\widehat w_{01},\widehat w_{11},\widehat w_{21})$ has been sent if there is a unique tuple of indices $(\widehat w_{01},\widehat w_{11},\widehat w_{21})$ such that $({u^n}(\widehat w_{01}),v_{1}^n(\widehat w_{01},\widehat w_{11},l),x_2^n(\widehat w_{01},\widehat w_{21}),{y_{1}^n}) \in A_{\varepsilon} ^n({P_{UV_1X_2Y_{1}}})$.
  \item Receiver 2 declares that the index pair of $(\widehat w_{02},\widehat w_{22})$ has been sent if there is a unique pair of indices $(\widehat w_{02},\widehat w_{22})$ such that $({u^n}(\widehat w_{02}),x_2^n(\widehat w_{02},\widehat w_{22}),{y_{2}^n}) \in A_{\varepsilon} ^n({P_{UX_2Y_{2}}})$.
\end{itemize}

Using joint decoding \cite{ElgamalKim}, it can be shown that the probability of error goes to zero as $n\rightarrow\infty$ if we choose:
{\small
\begin{align}
&R_{1} \le I({V_1};Y_{1}|{X_2},U)-I({V_1};Y_{2}|{X_2},U)\\
&{R_2} \le \min \{I({X_2};Y_{1}|{V_1},U),I({X_2};Y_{2}|U)\}\\
&{R_0} + R_{2} \le I(U,{X_2};Y_{2})\\
&R_{1} + R_{2} \le I(V_1,{X_2};Y_{1}|U) - I({V_1};Y_{2}|{X_2},U)\\
&{R_0} + R_{1} + R_{2} \le I(U,V_1,{X_2};Y_{1}) - I({V_1};Y_{2}|{X_2},U)
\end{align}}
\subsubsection{Equivocation computation}
{\small
\begin{align}\label{10}
H({W_1}|{Y_{2}^n}) &\ge H({W_1}|{Y_{2}^n},X_2^n,{U^n})\nonumber\\
&= H({W_1},{Y_{2}^n}|X_2^n,{U^n}) - H({Y_{2}^n}|X_2^n,{U^n})\nonumber\\
&= H({W_1},{Y_{2}^n},V_{1}^n|X_2^n,{U^n}) \nonumber\\
&\,\,\,- H(V_{1}^n|{W_1},{Y_{2}^n},X_2^n,{U^n}) - H({Y_{2}^n}|X_2^n,{U^n})\nonumber\\
&= H({W_1},V_{1}^n|X_2^n,{U^n})\nonumber\\
&\,\,\,+ H({Y_{2}^n}|{W_1},V_{1}^n,X_2^n,{U^n})\nonumber\\
&\,\,\,-H(V_{1}^n|{W_1},{Y_{2}^n},X_2^n,{U^n})- H({Y_{2}^n}|X_2^n,{U^n})\nonumber\\
&\ge H(V_{1}^n|X_2^n,{U^n}) + H({Y_{2}^n}|V_{1}^n,X_2^n,{U^n})\nonumber\\
&\,\,\,- H(V_{1}^n|{W_1},{Y_{2}^n},X_2^n,{U^n}) - H({Y_{2}^n}|X_2^n,{U^n})\nonumber\\
%\end{align}}therefore we have
%{\small
%\begin{align}\label{10}
&= H(V_{1}^n|{U^n}) - H(V_{1}^n|{W_1},{Y_{2}^n},X_2^n,{U^n})\nonumber\\
&\,\,\,- I(V_{1}^n;{Y_{2}^n}|X_2^n,{U^n})
\end{align}}where the last inequality is due to  the fact that $V_{1}^n$ is a function of ${W_1}$ and the last equality follows from the Markov chain $V_{1}^n - {U^n} - X_2^n$.
The first term in (\ref{10}) is given by:
\begin{equation}\label{avali}
H(V_{1}^n|{U^n}) = n\tilde R.
\end{equation}

We then show that $H(V_{1}^n|{W_1},{Y_{2}^n},X_2^n,{U^n}) \le n{\varepsilon _1}$, where as $n \to \infty $ then ${\varepsilon _1} \to 0$. Based on the Fano's inequality, we have:
\small{
\begin{equation}\label{bi}
  H(V_{1}^n|{W_1} = {w_1},{Y_{2}^n},X_2^n,{U^n}) \le 1 + {p_{e{w_1}}}(n\tilde R - n{R_1}) \equiv n{\varepsilon _1}\nonumber
\end{equation}}where $p_{e{w_1}}$ specifies the average of error probability of user~1 for decoding $v_{1}^n({w_0},{w_1},l)$ given ${W_1}={w_1}$ and ${W_0}={w_0}$. Hence
{\small
\begin{align}\label{dovomi}
&H(V_{1}^n|{W_1},{Y_{2}^n},X_2^n,{U^n})=\nonumber\\
&\sum\limits_{{w_1} \in {{\cal W}_1}} {p({W_1}={w_1})H(V_{1}^n|{W_1}={w_1},{Y_{2}^n},X_2^n,{U^n})}\le n{\varepsilon _1}.
\end{align}}

The last term in (\ref{10}) is bounded as:
\begin{equation}\label{sevomi}
I(V_{1}^n;{Y_{2}^n}|X_2^n,{U^n}) \le nI({V_1};Y_{2}|{X_2},U) + n{\varepsilon _2}
\end{equation}where as $n \to \infty $, then ${\varepsilon _2} \to 0$ \cite [Lemma 8]{Wyner}.
By replacing (\ref{avali})-(\ref{sevomi}) in equation (\ref{10}), and setting $\delta = {\varepsilon_1} + {\varepsilon_2}$, we have:
{\small
\begin{align}
H({W_1}|{Y_{2}^n}) &\ge n\tilde R - n{\varepsilon _1} - nI({V_{1}};Y_{2}|{X_2},U) - n{\varepsilon _2}\nonumber \\
&= n(\tilde R - I({V_{1}};Y_{2}|{X_2},U)) - n\delta \nonumber \\
&= n{R_1} - n\delta.
\end{align}}

This completes the proof of Theorem 2. \IEEEQED
\section{EXAMPLES}
In this section, we consider less noisy and Gaussian CMAC-CM.
\subsection{Less Noisy CMAC-CM}
We define 'less noisy' model as a CMAC-CM where $I(V_{2};Y_{1}) \geq I(V_{2};Y_{2})$ for all $p(v_{2},x_{2})$.

%%%%%%%%%%%%%%%%%%%%%%%%%%%%%%%%%%%%%%%%%%%%% Theorem 3 %%%%%%%%%%%%%%%%%%%%%%%%%%
\textbf{Theorem 3} (Outer Bound): The secrecy capacity region for the less noisy CMAC-CM is included in the set of rates satisfying
{\small
\begin{align}
\label{capacity R1}
&{R_1} \le I({V_1};Y_{1}|{V_2}) - I({V_1};Y_{2}|{V_2})\\
\label{capacity R2}
&{R_2} \le I({V_2};Y_{2})\\
\label{capacity R1+R2}
&{R_1} + {R_2} \le I({V_1},{V_2};Y_{1}) - I({V_1};Y_{2}|{V_2})\}
\end{align}}
for some joint distribution

$p(v_1,v_2)p(x_1|v_1)p(x_2|v_2)p(y_{1},y_{2}|x_1,x_2)$.

\textbf{Proof:} The proof is similar to Theorem 1. If we set $R_{0}=0$ and define $V_{1,i} = (Y_{2,i + 1}^n,{Y_{1}^{i - 1}},{W_1})$ and $V_{2,i} =(Y_{2,i + 1}^n,{Y_{1}^{i - 1}},{W_2})$, we can derive (\ref{capacity R1}) to (\ref{capacity R1+R2}).\IEEEQED

\textbf{Theorem 4} (Achievability): An inner bound on the secrecy capacity region is given by:
{\small
\begin{align}
\bigcup {\left\{ \begin{array}{l}
{R_1} \le I({V_1};Y_{1}|{V_2}) - I({V_1};Y_{2}|{V_2})\\
{R_2} \le I({V_2};Y_{2})\\
{R_1} + {R_2} \le I({V_1},{V_2};Y_{1}) - I({V_1};Y_{2}|{V_2})
\end{array} \right.}
\end{align}}where the union is taken over all joint distributions factorized as
$p(v_1)p(v_2)p(x_1|v_1)p(x_2|v_2)p(y_{1},y_{2}|x_1,x_2)$.

\textbf{Proof:} The proof follows from Theorem 2 by setting $U$ as a constant and by
replacing $X_2$ by $V_2$ where $V_2$ is obtained by passing $X_2$ through a DMC $p(v_2|x_2)$. Note that
{\small
\begin{align}
&I({V_2};Y_{1}|{V_1}) = H({V_2}|{V_1}) - H({V_2}|Y_{1},{V_1})\nonumber\\
&\mathop  = \limits^{(a)} H({V_2}) - H({V_2}|Y_{1},{V_1})= I({V_2};Y_{1},{V_1})\nonumber\\
&\ge I({V_2};Y_{1})\mathop  \ge \limits^{(b)} I({V_2};Y_{2})\nonumber
\end{align}}where $(a)$ happens since $V_1$ and $V_2$ are independent and $(b)$ results from considering less noisy model. Therefore, we have $\min\{I({V_2};Y_{1}|{V_1}),I({V_2};Y_{2})\}=I({V_2};Y_{2})$.\IEEEQED

Note that although the outer and inner bounds for this model has the same characteristic, they are over two different joint distributions.
%%%%%%%%%%%%%%%%%%%%%%%%%%%%%%%%%%%%%%%%%%%%%%%%%%%%%%%%%%%%%%%%%%%%%%%%%%%
\subsection{Gaussian CMAC-CM}
\begin{figure}%*}[ht]
%  \centering
  \includegraphics[width=8.87cm]{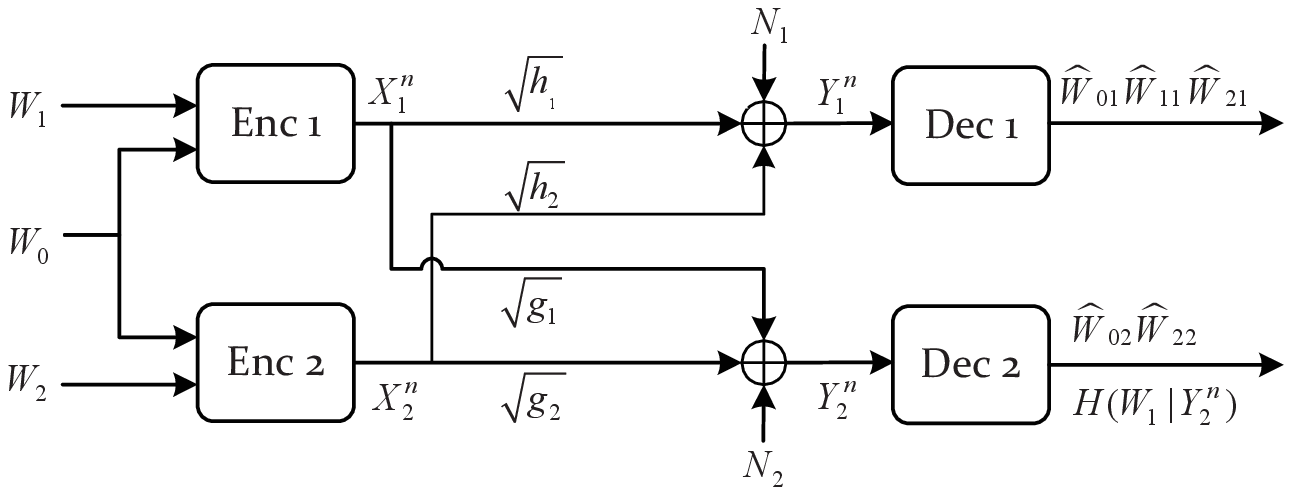}% asli 12.50cm
  \caption{\small{Gaussian compound MAC with confidential messages.}}
  \label{figIV}
\end{figure}%*}
In this section, we consider Gaussian CMAC-CM as depicted in Fig. \ref{figIV}, and extend the achievable rate region of discrete memoryless CMAC-CM to the Gaussian case. Relationships between the inputs and outputs of the channel are given by
{\small
\begin{align}
\label{Y_1}
Y_{1} = \sqrt {{h_1}} X_{1} + \sqrt {{h_2}} X_{2} + {N_{1}}\\
\label{Y_2}
Y_{2} = \sqrt {{g_1}} X_{1} + \sqrt {{g_2}} X_{2} + {N_{2}}
\end{align}}where $h_i, g_i$ (for $i = 1,2$) are known channel gains as shown in Fig. \ref{figIV}. The noise terms $N_{1}$ and $N_{2}$ are independent zero-mean unit-variance complex Gaussian random variables and independent of the real random variables $X_1$, $X_2$. The inputs of the channel are also subject to an average power constraint
{\small
\begin{equation}\label{power}
\frac{1}{n}\sum\limits_{i = 1}^n {E[X_{k,i}^2]}  \le {P_k}\,\,\,\,\,\,\,k = 1,2.
\end{equation}}

Let $V_1, V_2, X_1$, and $X_2$ be jointly Gaussian random variables with
\begin{align}
\label{V_1}
{V_1} &= \sqrt {{P_{{U_1}}}} U + \sqrt {{P_{U'}}} U',\,\,%\\ \nonumber
{V_2} = \sqrt {{P_{{U_2}}}} U + \sqrt {{P_{U''}}} U''\\
\label{X_1vaX_2}
{V_1} &\buildrel \Delta \over = {X_1},\,\,{V_2} \buildrel \Delta \over = {X_2}
\end{align}
where $U, U'$ and ${U''}$ are independent Gaussian zero mean unit variance random variables. The terms $P_{U_1}, P_{U_2}, P_{U'}$, and $P_{U''}$ denote the corresponding power allocation, where
\begin{equation}
\label{P_1}
{P_1} = {P_{{U_1}}} + {P_{U'}},\,\,%\\ \nonumber
{P_2} = {P_{{U_2}}} + {P_{U''}}.
\end{equation}

Following the achievability proof for the discrete memoryless channel, we obtain the following result for the Gaussian CMAC-CM.

\textbf{Theorem 5:} An inner bound on the secrecy capacity region of Gaussian CMAC-CM is:
{\small
\begin{equation}\label{GAchievable}
\bigcup \left\{ \begin{array}{l}
%\[\bigcup {\left\{ \begin{array}{l}
{R_1} \le C({h_1}{P_{U'}}) - C({g_1}{P_{U'}})\\
{R_2} \le \min \{ C({h_2}{P_{U''}}) , C(\frac{{{g_2}{P_{U''}}}}{{1 + {g_1}{P_{U'}}}})\}\\
{R_0} + {R_2} \le C(\frac{{{g_1}{P_{{U_1}}} + {g_2}{P_2} + 2\sqrt {{g_1}{g_2}{P_{{U_1}}}{P_{{U_2}}}} }}{{1 + {g_1}{P_{U'}}}})\\
{R_1} + {R_2} \le C({h_1}{P_{U'}} + {h_2}{P_{U''}}) - C({g_1}{P_{U'}})\\
{R_0} + {R_1} + {R_2} \le C({h_1}{P_1} + {h_2}{P_2} + 2\sqrt{{h_1}{h_2}{P_{{U_1}}}{P_{{U_2}}}} )\\
\,\,\,\,\,\,\,\,\,\,\,\,\,\,\,\,\,\,\,\,\,\,\,\,\,\,\,\,\,\,\,\,\,\,\,\,\,\,\,\,\,\,\
- C({g_1}{P_{U'}})
%\end{array} \right.} \]}
\end{array} \right.
\end{equation}}where $C(x) = (1/2)\log (1 + x)$ and the union is taken over all $0 \le {P_{{U_1}}} + {P_{U'}} \le {P_1}$ and $0 \le {P_{{U_2}}} + {P_{U''}} \le {P_2}$.

\textbf{Proof:} Following from Theorem 2, by choosing random variables the same as (\ref{Y_1})-(\ref{P_1}) the region in (\ref{GAchievable}) is derived.
\begin{figure}%*}[ht]
%  \centering
  \includegraphics[width=9.50cm]{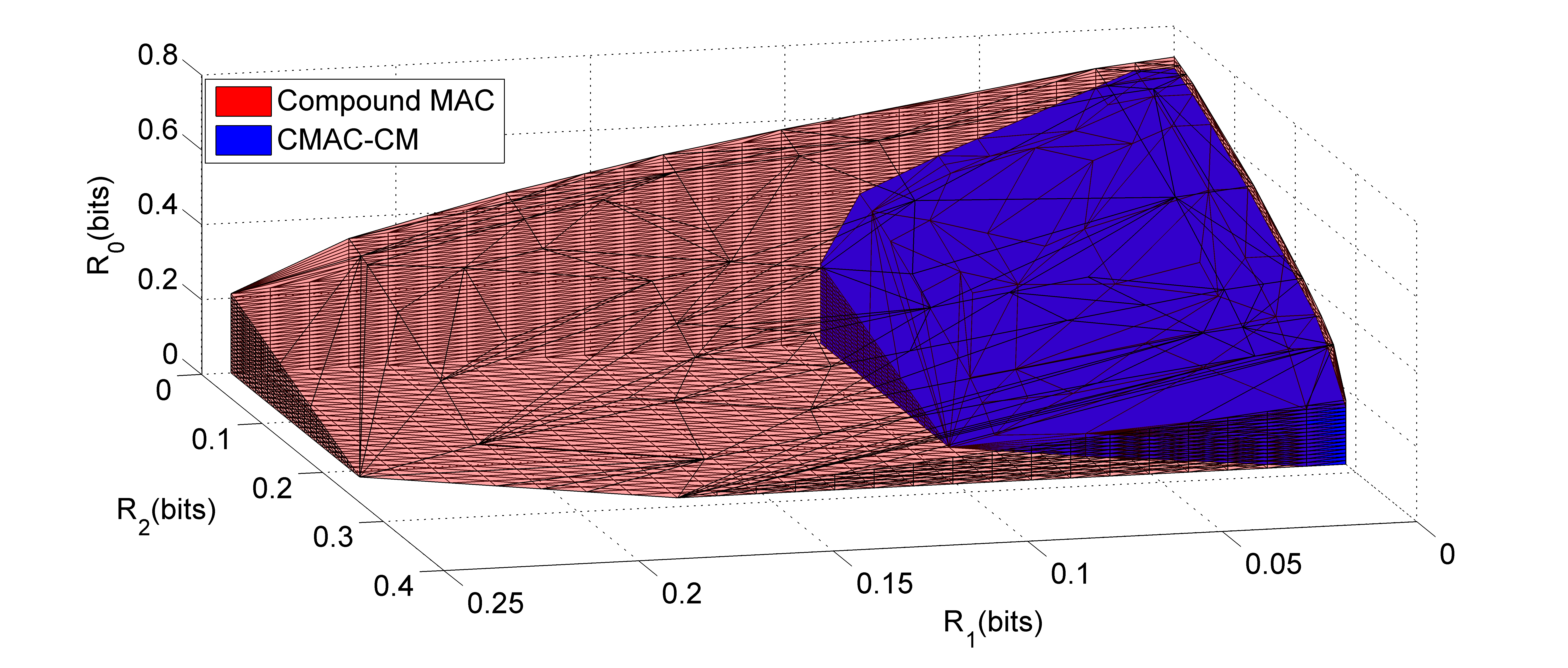}% asli 12.50cm
  \caption{\small{Achievable rate region of Gaussian CMAC-CM and the capacity region of Compound MAC for $P_1=P_2=1$, $h_1=h_2=0.6$, $g_1=0.4$ and $g_2=0.5$.}}
  \label{figII}
\end{figure}%*}

\textbf{Theorem 6:} The capacity region of Compound Gaussian MAC is given by:
{\small
\begin{equation}\label{capacityCMAC}
\bigcup \left\{ \begin{array}{l}
0 \le {R_1} \le \min \{ C({h_1}{P_{U'}}),C({g_1}{P_{U'}})\} \\
0 \le {R_2} \le \min \{ C({h_2}{P_{U''}}),C({g_2}{P_{U''}})\} \\
0 \le {R_1} + {R_2} \le \min \{ C({h_1}{P_{U'}} + {h_2}{P_{U''}})\\
\,\,\,\,\,\,\,\,\,\,\,\,\,\,\,\,\,\,\,\,\,\,\,\,\,\,\,\,\,\,\,\,\,\,\,\,\,\,\,\,\,\,\,\,\,\,\,\,\
,C({g_1}{P_{U'}} + {g_2}{P_{U''}})\} \\
0 \le {R_0} + {R_1} + {R_2} \le \\
\,\,\,\,\,\,\,\,\,\,\,\,\,\,\,\,\,\min \{ C({h_1}{P_1} + {h_2}{P_2} + 2\sqrt {{h_1}{h_2}{P_{{U_1}}}{P_{{U_2}}}}),\\
\,\,\,\,\,\,\,\,\,\,\,\,\,\,\,\,\,\,\,\,\,\,\,\,\,\,\,\,\
C({g_1}{P_1} + {g_2}{P_2} + 2\sqrt {{g_1}{g_2}{P_{{U_1}}}{P_{{U_2}}}} )\}
\end{array} \right.
\end{equation}}where $C(x) = (1/2)\log (1 + x)$ and the union is taken over all $0 \le {P_{{U_1}}} + {P_{U'}} \le {P_1}$ and $0 \le {P_{{U_2}}} + {P_{U''}} \le {P_2}$.

\textbf{Proof:} Considering Propositions 6.1 and 6.2 in \cite{simeone} which are outer and inner bounds on the capacity region of compound MAC with conferencing links, then ignoring conferencing links in that model (i.e, setting $C_{12}=C_{21}=0$ in \cite{simeone}), and by considering (\ref{Y_1}) to (\ref{P_1}), the region in (\ref{capacityCMAC}) is derived.\IEEEQED

As an example, for the values $P_{1}=P_{2}=1$, $h_{1}=h_{2}=.6$, $g_{1}=.4$ and $g_{2}=.5$ the achievable rate region in Theorem 5 and the capacity region in Theorem 6 are depicted in Fig. \ref{figII}. As it can be seen in Fig. \ref{figII}, for this scenario the achievable rate $R_{0}$ (rate of $W_{0}$ which is decoded by both receivers) is nearly the same for both CMAC-CM and compound MAC models. Moreover, the achievable rate $R_{2}$ (rate of $W_{2}$ which is also decoded by both receivers) is equal for both CMAC-CM and compound MAC models. Moreover, as it can be seen from Fig. \ref{figII}, for these channel gain parameters the achievable rate $R_1$ for CMAC-CM is less than that for Compound MAC due to secrecy constraint for decoding message $W_1$.

\begin{figure}%*}[ht]
%  \centering
  \includegraphics[width=9.50cm]{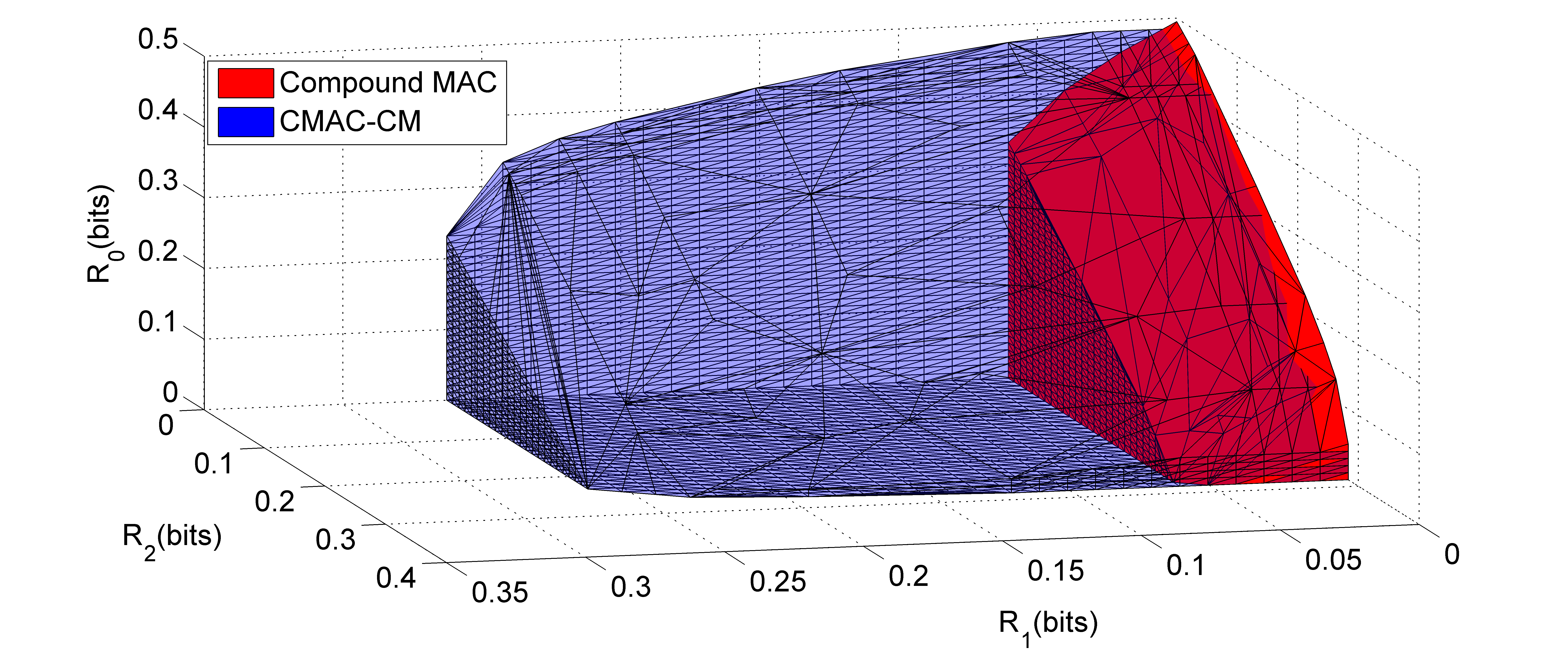}% asli 12.50cm
  \caption{\small{Achievable rate region of Gaussian CMAC-CM and the capacity region of Compound MAC $P_1=P_2=1$, $h_1=h_2=0.6$, $g_1=0.1$ and $g_2=0.5$.}}
  \label{figIII}
\end{figure}

According to (\ref{capacityCMAC}), it is clear that if the channel gain between transmitter 1 and receiver 2 (i.e., $g_1$) decreases, the capacity region of compound MAC does not increase (i.e., the capacity region may remain as before or decreases). On the other hand, there exist scenarios for CMAC-CM (refer to (\ref{GAchievable})) for which decreasing $g_1$ results in increasing its achievable rate region. For comparison, assume changing $g_1=0.4$ to $g_1=0.1$ in the above example. As it can be seen in Fig. \ref{figIII}, the achievable rate region of CMAC-CM is larger than the capacity region of compound MAC model for the new parameters. This can be interpreted as follows: transmitted signal of transmitter 1 is extremely attenuated at the receiver 2. So, for this case, not forcing the receiver 2 to decode $W_1$ (keeping $W_1$ as a secret for receiver 2) can increase the achievable rate region.
%*}

Also, it should be noted that according to (\ref{GAchievable}), for the scenarios where $g_1$ equals or greater than $h_1$, the achievable rate $R_1$ is zero. This happens since for these scenarios the channel gain between transmitter~1 and receiver~2 (i.e., illegal user in terms of message $W_1$) is equal or better than the channel gain between transmitter~1 and receiver~1 (i.e., legitimate user in terms of message $W_1$), and this makes zero $R_1$.
\section{Conclusions}
In this paper, we have studied the secrecy capacity region of the Compound MAC with one Confidential Message (CMAC-CM). We have obtained inner and outer bounds on the secrecy capacity for the general CMAC-CM. We have further studied less noisy and Gaussian CMAC-CM. Providing numerical examples for the Gaussian case we have shown that there are scenarios for which keeping secret some messages can increase the rate region compared to being required to communicate those messages reliably to the other receiver.
\bibliographystyle{IEEEtran}
\bibliography{IEEEexample,mybibfile}
\end{document}